\documentclass[12pt,preprint]{aastex}

\shorttitle{Detection of an AGN in a $z=3.09$ Ly$\alpha$ halo}
\shortauthors{Basu-Zych \& Scharf}

\begin{document}


\title{X-Ray Detection of an Obscured Active Galactic Nucleus in a $z=3.09$ Radio-quiet Ly$\alpha$ Nebula}


\author{Antara Basu-Zych}
\affil{Department of Astronomy, Columbia University, Mail Code 5246, 550 West 120th Street, New York, NY 10027, USA}
\email{antara@astro.columbia.edu}

\author{Caleb Scharf}
\affil{Columbia Astrophysics Laboratory, Columbia University, MC 5247, 550 West 120th Street, New York, NY 10027, USA}\email{caleb@astro.columbia.edu}


\begin{abstract}
We present evidence for a highly obscured X-ray source in one of two giant Ly$\alpha$
emission nebulae in the $z=3.09$ proto-cluster region SSA22. Neither Ly$\alpha$ nebula
is associated with significant radio emission. While one has a significant submillimeter 
detection and is undetected in the X-ray, the other is a factor of 2-10 times less submillimeter
bright and appears to contain a hard-band X-ray source. We discuss our analysis and
techniques for assessing the X-ray properties of this source and suggest that
we have detected an embedded AGN source in one of these nebulae which may be
at least partially responsible for exciting the Ly$\alpha$ emission through a
mechanism that is essentially decoupled from the radio, submillimeter, or optical luminosities.
We also present an
upper limit on the mean X-ray emission from 10 other extended Ly$\alpha$ 
objects in the SSA22 region.

\end{abstract}


\keywords{galaxies:high-redshift---galaxies:evolution---galaxies:formation---X-rays:galaxies}


\section{Introduction}

The standard paradigm of structure formation in a universe dominated by
cold dark matter suggests that the high-density regions of
the high-redshift universe that later collapse into clusters should also
be host to the progenitors of massive galaxies. Identifying both
proto-cluster regions and proto-elliptical galaxies therefore offers
an opportunity to directly witness the astrophysical phenomena
that likely set many of the characteristics of clusters and
cluster galaxies seen today.

The identification of protocluster regions at $z>2$ remains 
a challenging task. Among the approaches used are the localization of
spatial excesses of objects such as Lyman break galaxies \citep{ste98}, Lyman emission line
objects (e.g., \citet{hu98}), and submillimeter sources (e.g., \citet{ivi00}),
and the apparent correlation of QSO absorption features along adjacent
lines of sight \citep{fra96}. A slightly different approach involves the use
of signpost objects, such as luminous radio galaxies, which
are expected to form in the highest density regions at high redshift \citep{ste03}.
All of the above methods have yielded positive results with varying degrees
of success.

In a survey of continuum-selected Lyman break galaxies \citet{ste98,ste00}
have found evidence for a highly significant overdensity of objects at $z=3.09$
(a factor of $\sim 5$ higher number density than
the average equivalent volume at this redshift). In subsequent narrowband
Ly$\alpha$ imaging of this field, two very bright, large, diffuse Ly$\alpha$-emitting
nebulae (or ``halos'' or ``blobs'') were discovered. Named ``blob 1'' (B1) and ``blob 2'' (B2)
by \citet{ste00} these objects have sizes of 17'' and 15'' (B1, B2), Ly$\alpha$
luminosities of $10^{44}$  ergs s$^{-1}$(B1) and $9\times 10^{43}$ ergs s$^{-1}$ (B2), and rest
equivalent widths of {\em at least} $370$ \AA. 
Both blobs have radio continuum fluxes at 1.4 GHz  at least  3 orders of magnitude
fainter than those of typical luminous high-redshift radio galaxies (B1 has a tentative
radio identification with a 44 $\mu$Jy [1.4 GHz] source; \citet{cha04}). 
High-$z$ Ly$\alpha$ halos certainly exist with very powerful radio sources and jets 
(e.g., \citet{bre03,reu03}); however these have been
preselected by their radio luminosity, and so it is not clear whether
low radio activity is common or not in such systems. submillimeter observations of the
blobs have detected sources that (given the $\sim 15''$ half-power beamwidth at 850 $\mu$m)
are potentially counterparts to B1 and B2 \citep{cha01,cha04}. However, while the B1 counterpart was
detected as a $20.1\pm3.3$ mJy source at 850 $\mu$m, for B2 the submillimeter emission was significantly
less at $3.3\pm1.2$ mJy. Thus, the B1 submillimeter counterpart is one of the most luminous
submillimeter high-$z$ sources known, while the B2 counterpart is of low significance.

In recent follow-up work on B1, \citet{cha04} have further identified
several optical continuum components (to a limit $R\sim 28.6$) that
may be associated with the core of the submillimeter source, and that appear
within, and are potentially associated with, the 17$''$ ($\sim 140$
kpc) Ly$\alpha$ blob. They also obtain a marginal detection of a
CO(4-3) molecular line and estimate a molecular gas mass of $2\times
10^{10}$M$_{\odot}$. Furthermore, \citet{cha04} find no significant X-ray emission
associated with B1 using deep {\sl Chandra} data, which we further
analyze below.

The current physical interpretation of the B1 structure is that it is
consistent with an undetected, dust-obscured active galactic nucleus (AGN), together with a starburst region embedded at
the  core of  the  Ly$\alpha$  nebula. Indeed, \citet{cha04}
speculate that a highly obscured  AGN could still be responsible for a
jet  structure inclined  well  away from  our  line of sight that is
inducing star  formation with the  observed spatial morphology. Ionizing
flux from the AGN in such a geometry could then also assist in lighting up
the Ly$\alpha$ cloud. The only obvious difficulty  with this interpretation  
is the apparent  absence of extended radio structure. We will return to these 
scenarios in the Discussion below.

\section{Data and Analysis}

We analyze archival data, taken with the ACIS-S detector onboard {\it
Chandra} on 2001 July 10. The 78 ks observation was centered at
$\alpha_{2000}=22^{h}17^{m}32.40^{s},
\delta_{2000}=+00^{\circ}13'09.9''$, well-situated for studying the
Ly$\alpha$ blobs detected by \citet{ste00} located at
$\alpha_{2000}=22^{h}17^{m}25.7^{s},
\delta_{2000}=+00^{\circ}12'49.6''$ (B1) and
$\alpha_{2000}=22^{h}17^{m}39.0^{s},
\delta_{2000}=+00^{\circ}13'30.1''$ (B2).
Previously, these data have been analyzed by \citet{cha04} (for B1) and 
by \citet{bau02}. 

We follow the standard
procedure of eliminating high background by excluding times where the
counts exceeded quiescent levels. This observation was
fortunate to have not occurred during any flares, and no time
is excluded. Next we create the instrument map, using the CIAO
{\tt mkinstmap} command, for the ccd chip containing our data: the
back-illuminated ACIS-S3. As described below, final photon counts
for the detected source are very low; consequently, we determine
monochromatic exposure maps in two bands (spatially binned by 4 to $2''$ pixels),
with the energy chosen to
match the most probable photon energy in the observed counts in each band, namely,
hard (2-8 keV) : 3.0 keV; and soft
(0.3-2 keV): 0.8 keV. The total band map (0.3-8 keV) was formed by
averaging these maps. Finally, image maps with units of
photon cm$^{-2}$ s$^{-1}$ pixel$^{-1}$ are constructed.

The smoothed, 0.3-8 keV X-ray image of the B2 region is compared to
 the Ly$\alpha$ image in Figure 1. There is visual evidence for
 correlation of the X-ray image with the Ly$\alpha$ image and possibly
even with the extended Ly$\alpha$ features.
 The region of B1 is not apparent in any of the
 soft, hard (Figure 2) or total X-ray images, and the soft X-ray image
 does not seem to contain evidence for B2 (Figure 3). To quantify
 these observations we use three separate techniques. First we apply
 the {CIAO} {\sl wavdetect} algorithm which convolves the image with
 a ``Mexican hat'' wavelet kernel to identify sources. In both the
 hard image and the total image {\sl wavdetect} finds sources coincident
 with the Ly$\alpha$ position for B2, with a signal-to-noise ratio
 determination of 2.7 (2-8 keV) and 3.6 (0.3-8 keV); however,
 {\sl wavdetect}  does not find any sources in the soft image of B2 or
 for B1 in any of the bands. Our second method serves to verify these
 results manually. In our manual test, we integrate the counts within
 $8''$ of the source. Taking {\sl Chandra}'s point-spread function (PSF) into account, an
 $8''$ radius aperture conservatively contains $\sim$95\% of the flux
 of a point source at 8 keV and more at lower energies. The size of this
hard PSF and low signal-to-noise ratio complicates the interpretation of
any possible extended nature of the B2 feature, and it is 
conceivable that the emission is not due to a single point source.
 We determine a background count by choosing an annulus surrounding our
 source with an inner radius of $9''$ and outer radius of $25''$. Our
 calculations of the signal-to-noise ratio are consistent with the {\sl wavdetect}
 results.

A third test for the significance of the B2 X-ray counterpart is to
determine the probability that the detection is a chance coincidence of
X-ray emission at the B2 location. 
By placing 1000 random $8''$ apertures throughout
the field and integrating the hard and total band flux within each aperture to obtain the
probability distribution of counts we determine
the likelihood of the total count exceeding that obtained in
the B2 region. From this Monte Carlo analysis we determine that our detection is
98\% significant for the total band image, and 99\% significant for the
hard band image. The same test performed on the soft band yields a 90\% significance, 
which corresponds to less than a 2 $\sigma$ detection for Poissonian distributions.
 We therefore conclude that a hard X-ray source exists at this location and
therefore may be associated with the Ly$\alpha$ emission,
B2. However, we concede that the source is extremely faint and without
the agreement of these separate tests and previous knowledge of the
Ly$\alpha$ source, one might not consider this detection relevant;
such was the conclusion reached by \citet{bau02}.

The further possibility that the X-ray source is a chance alignment
 of an unassociated AGN with the Ly$\alpha$ blob has been investigated. Using the
 study by \citet{bau04} of X-ray number counts in the Chandra Deep
 Field, the number of sources per square degree with flux greater
 than or equal to the B2 source implies only $\simeq 0.016$ sources
 per $8''$ aperture. We are therefore 98\% confident that the X-ray source and
 the Ly$\alpha$ source are indeed related.

The background-subtracted 2-8 keV count rate within an $8''$ aperture centered in B2
is $1.3 (\pm 1.1) \times 10^{-4}$ counts $s^{-1}$ ($24 \pm 20$ counts). As discussed, the rate in the soft band 
is indistinguishable from  the
background. The hardness ratio ($H-S/H+S$) for B2 is therefore consistent with
unity. In order to assess the possible degree of absorption responsible for
this spectral hardening we have assumed a generic, intrinsic power-law X-ray
spectrum of photon index $1.4$, and a measured Galactic foreground
absorbing column of $n_{H}=4.8\times 10^{20}$ cm$^{-2}$. We then use XSPEC
to determine the intrinsic luminosity and additional absorbing column (assuming solar abundances)
necessary to reproduce the observed {\sl Chandra} upper limit to the soft-band
count rate and the hard-band count rate. Table 1 summarizes the resulting
luminosity estimates (absorbed and unabsorbed) corresponding to observed and
emitted energy bands. The required absorber column
density is $\gtrsim 9\times 10^{22}$ cm$^{-2}$. 
Assuming this best fit model we then estimate the observed 2-8 keV flux to be 2.2$\pm 2.0 \times 10^{-15}$
 ergs cm$^{-2}$ s$^{-1}$, the 0.3-2 keV 3 $\sigma$ upper limit flux of 5.4$\times 10^{-16}$ ergs cm$^{-2}$ s$^{-1}$, 
 and a total 0.3-8 keV flux of 2.7$\pm 2.0 \times 10^{-15}$ ergs cm$^{-2}$ s$^{-1}$. 

If the X-ray source is indeed point-like, the estimate (Table 1) 
for an unabsorbed X-ray luminosity
 of $\sim 10^{44}$ erg $s^{-1}$ is not inconsistent with the B2 region
 harboring a supermassive black hole (SMBH), with a very high local absorbing column. 
Such a column is physically unlikely for an extended source, although the
possibility remains for more than one point source associated with B2.

\subsection{Upper limits on Ly$\alpha$ blob counterparts}

\citet{mat04} have published a catalog of 35 Ly$\alpha$ blob
candidates across the broader SSA22 field. We have examined the 12
candidates that are covered by the {\sl Chandra} ACIS-S3 chip in the
data considered here. {\sl Wavdetect} picked out three sources that
roughly matched the coordinates from the catalog, including
B2. However none of these six had signal-to-noise ratio values greater than
2, except B2. Excluding B1 and B2, the remaining 10 blobs have no
obvious X-ray counterparts. The published positions of the blobs are
given to a precision of only $6''$, so it is impossible to rule out
all counterparts, however, there are only two cases where an X-ray
source lies on the edge of a $6''$ radius region.  We have therefore
stacked the X-ray data extracted in $8''$ radius regions around these
10 Ly$\alpha$ sources. The Monte Carlo technique described in the
previous section evaluated an 85\% significance for the hard stacked
data and only a 66\% significance for the soft stacked data. We
determined the $3 \sigma$ upper limit to a mean X-ray count rate to be
$\sim 8\times 10^{-5}$ counts $s^{-1}$ (in the observed 0.3-8 keV). This
corresponds to an upper limit on the unabsorbed X-ray flux or
luminosity (assuming $z=3.09$) of $\sim 10^{-15}$ ergs cm$^{-2}$
s$^{-1}$ or 7$\times 10^{43}$ ergs $s^{-1}$ respectively.

\section{Summary and Discussion}

The evidence presented above for the presence of obscured AGN emission
of $L_x\sim 10^{44}$ erg s$^{-1}$ embedded in the Ly$\alpha$
cloud/B2 of SSA22 provides the first direct indication of AGN
activity in this possibly unusual, radio-quiet, object.  This
detection provides significant clues as to what is going on in its
``companion'' Ly$\alpha$ cloud, B1, and possibly to the generation of
the Ly$\alpha$ cloud emission itself.

It is useful to compare and contrast B1 and B2. These objects have comparable Ly$\alpha$
luminosities, both are detected in the submillimeter, while B2 is a factor of $\sim 10$ less luminous,
and while B2 has an X-ray detectable, albeit highly obscured, AGN, B1 does not to the current
sensitivity limits.

Various scenarios can be constructed from these observations. For
example, {\em if} the observed submillimeter luminosity is directly
indicative of the mass of dust in a system (assuming similar
temperatures) then it is tempting to conclude that B2 is less
enshrouded than B1 and is therefore more transparent to high-energy
X-rays (8-33 keV rest frame). While this may be correct, it is also
possible that the obscuration of the AGN (in B2 for certain) is due to
a much smaller scale molecular/dust torus surrounding the SMBH. The similarity in Ly$\alpha$ luminosity is
intriguing. If the ionizing source for the Ly$\alpha$ cloud were
directly connected with that responsible for heating the submillimeter dust
(i.e. starburst activity) then some amount of coincidental tuning
would be needed to produce a similar level of emission between B1 and
B2.

Finally, we  offer a different, albeit speculative, mechanism that
could explain many,  if  not   all,  of  the  above  features  of
the  B1  and  B2 systems. Recently, evidence has  been found for
extended X-ray emission around  high-$z$  radio-loud   galaxies  due
to  the  inverse Compton scattering of  cosmic microwave background (CMB)
photons by  a population of relativistic particles originating from
the central SMBH in these objects \citep{fab03,sch03}.  Owing to  the
$(1+z)^{4}$  growth  of  the  CMB  energy density,  this  emission  is
increasingly luminous at high-$z$, and  can readily exceed that of the
central AGN \citep{sch02,sch03}. In at  least one case \citep{fab03},
much of the X-ray emission appears  to be due to an {\em  older}
relativistic population of much greater spatial extent than the
currently detected radio synchrotron emission. The presence of a
spatially extended, highly luminous, X-ray emission offers a ready
mechanism for photo-ionization of cooler gas. Indeed, \citet{sch03}
speculate that it alone could sustain a $\sim 100$ kpc Ly$\alpha$
emission cloud, while not over-ionizing observed species such as those
of oxygen. If B2 (and perhaps B1) contains an AGN, then it is plausible
that a population of older relativistic particles exist on scales of
$\sim 100$ kpc, unseen via radio synchrotron since the typical
emission frequencies will be less than 100 MHz (rest frame) but still
actively upscattering CMB photons to the X-ray. Some of the
``triangular shaped'' or hourglass, morphology of the Ly$\alpha$
clouds reported in B1 \citep{cha04} and seen in B2 \citep{ste00}
matches extremely well the morphologies seen in the X-ray surrounding
3C 294 ($z=1.89$; \citet{fab03}) and in {\em both} the X-ray and
Ly$\alpha$ seen in 4C 41.17 ($z=3.79$; \citet{sch03}).  The relatively
low X-ray surface brightness of such emission ($\sim 8\times 10^{-17}$
ergs s$^{-1}$ cm$^{-2}$ arcsec$^{-2}$) would indicate that it can
readily escape direct detection even in a 78 ks ACIS-S
observation - although as discussed in Section 2, there is some suggestion that
the  B2 X-ray emission could be extended. The phenomenon described above offers an
interpretation if this is the case.

In such a picture, the relative submillimeter, radio, and optical luminosities
associated with the Ly$\alpha$ clouds are essentially decoupled from
the mechanism responsible for exciting their fluorescent emission - which is
driven by a periodic injection of relativistic particles (as seems
to be a standard AGN property), and the enhanced energy density of
the CMB (a factor of 280 at $z=3.09$).

\acknowledgments

This work was made possible through the use of the {\sl Chandra} archive, which is
part of the {\sl Chandra X-Ray Observatory} Science Center (CXC) and is operated for NASA by the Smithsonian Astrophysical Observatory (SAO).
C.S. acknowledges the support of NASA/{\sl Chandra} grant SAO G03-4158A and 
the Columbia Astrophysics Laboratory. We would also like to thank the anonymous referee for valuable comments and suggestions.

\clearpage

\clearpage
\begin{deluxetable}{cccc} 
\tablecolumns{4} 
\tablewidth{0pc} 
\tablecaption{Estimated absorbed and unabsorbed X-ray model luminosities in {\em rest frame} energy ranges ($E_{em}$) 
for B2 (corresponding observed energy ranges given under $E_{obs}$.)} 
\tablehead{ 
\colhead{$E_{em}$} & \colhead{$E_{obs}$}   & \colhead{$L_{abs}$\tablenotemark{a}}    & \colhead{$L_{unabs}$} \\
\colhead{(keV)}    & \colhead{(keV)}   & \colhead{($10^{44}$ ergs/s)}    & \colhead{($10^{44}$ ergs/s)} \\
}
\startdata 
2$-$8 & 0.5$-$2 & 0.46 \tablenotemark{*}& 0.87\tablenotemark{*} \\
0.3$-$2 & 0.08$-$0.5 & 0.00047\tablenotemark{*} & 0.42\tablenotemark{*}\\
0.3$-$8 & 0.08$-$2 & 0.46\tablenotemark{*} & 1.3\tablenotemark{*} \\
1.2$-$8 & 0.3$-$2 & 0.46\tablenotemark{*} & 1.0\tablenotemark{*} \\
8$-$33 & 2$-$8 & 1.9$\pm 1.6$ & 2.0$\pm 1.7$ \\
1.2$-$33 & 0.3$-$8 & 2.3$\pm 1.7$ & 3.1$\pm 2.2$ \\

\enddata 
\tablenotetext{*} {These are upper limits.}
\tablenotetext{a} {Luminosities were calculated using the following cosmology: H$_{0}$= 71 km s$^{-1}$Mpc$^{-1}$,
 $\Omega_{M}$=0.270, $\Omega_{\nu}$= 0.730}
\end{deluxetable} 






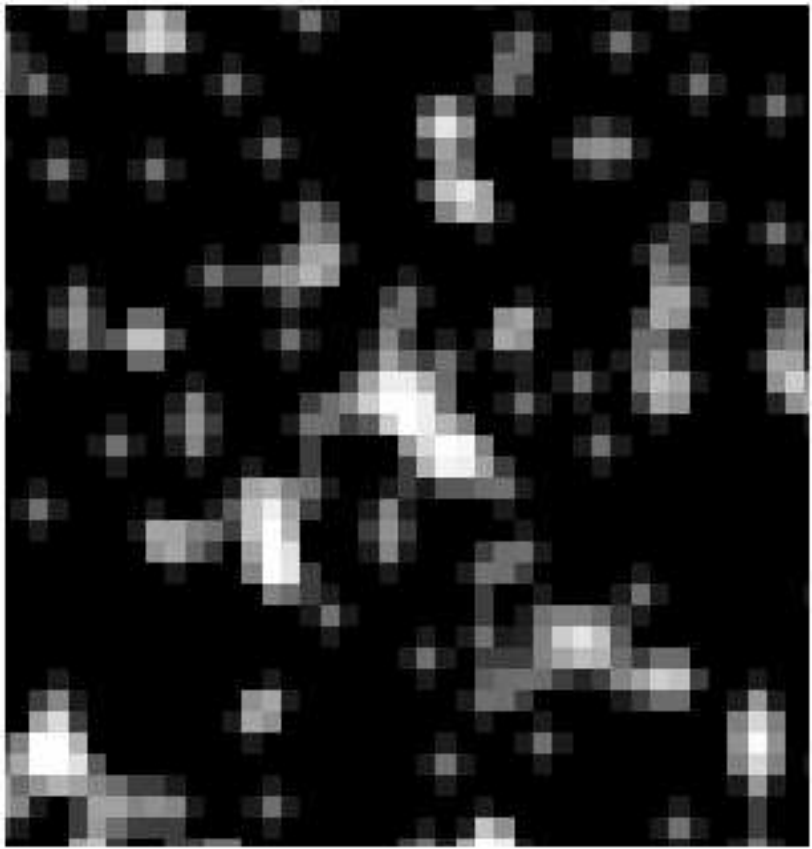
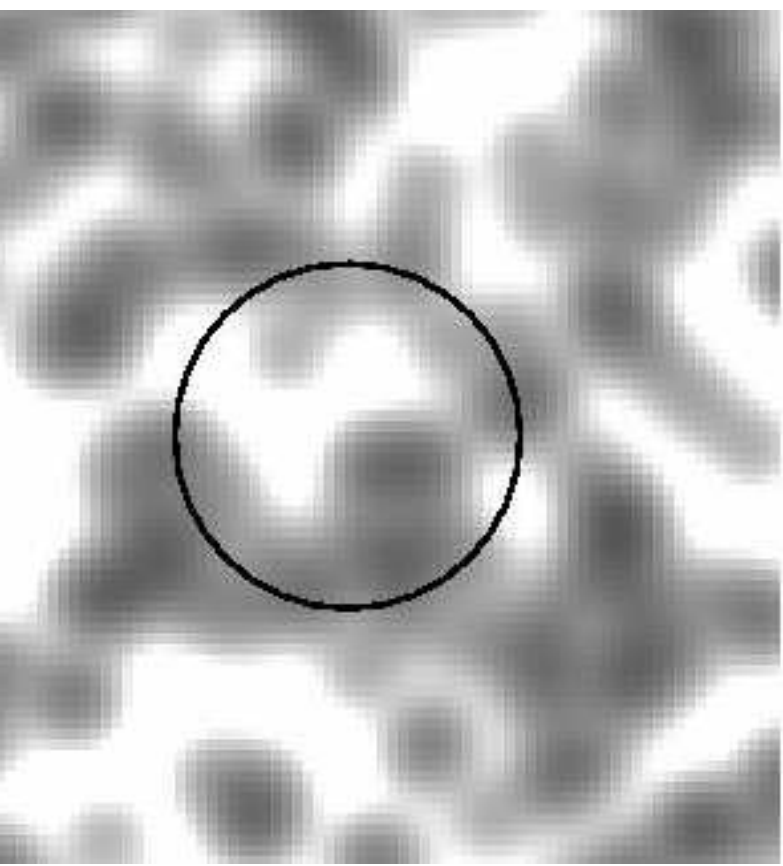
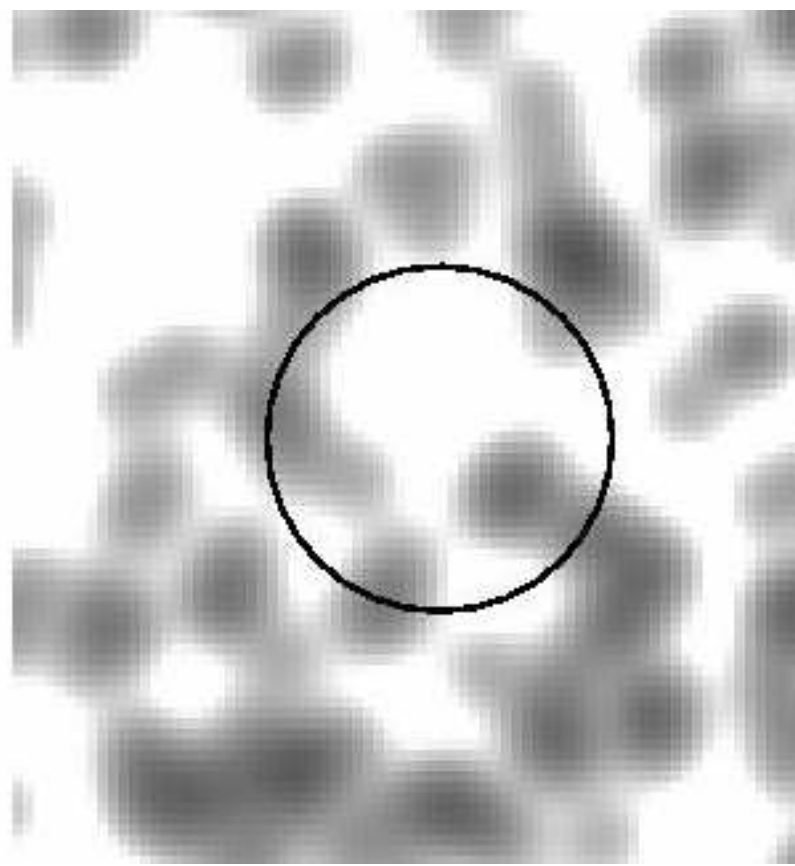
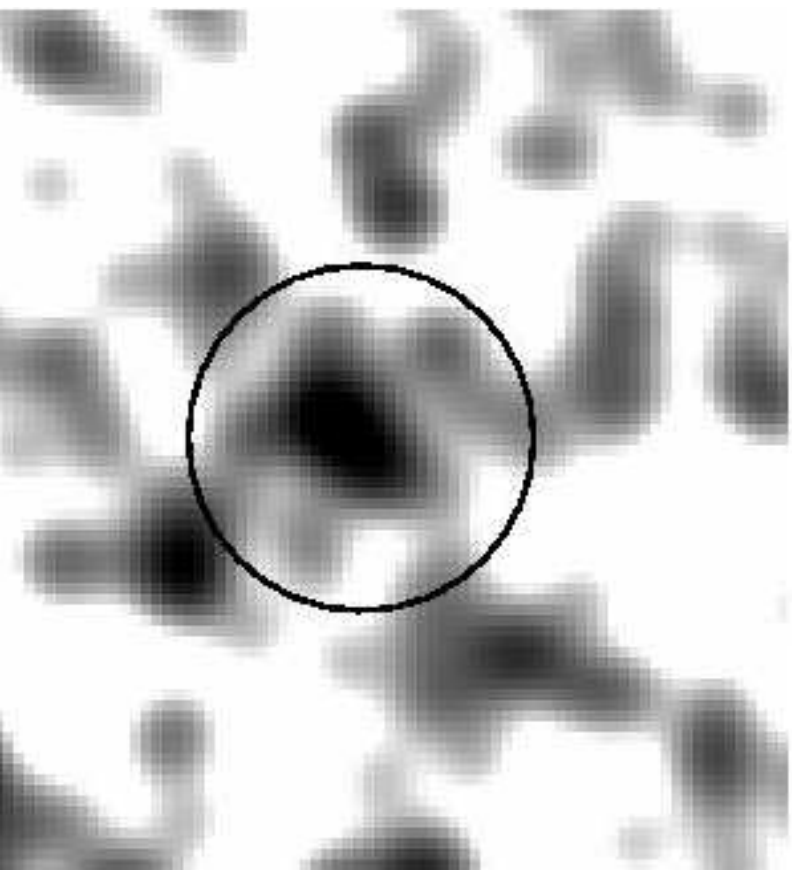
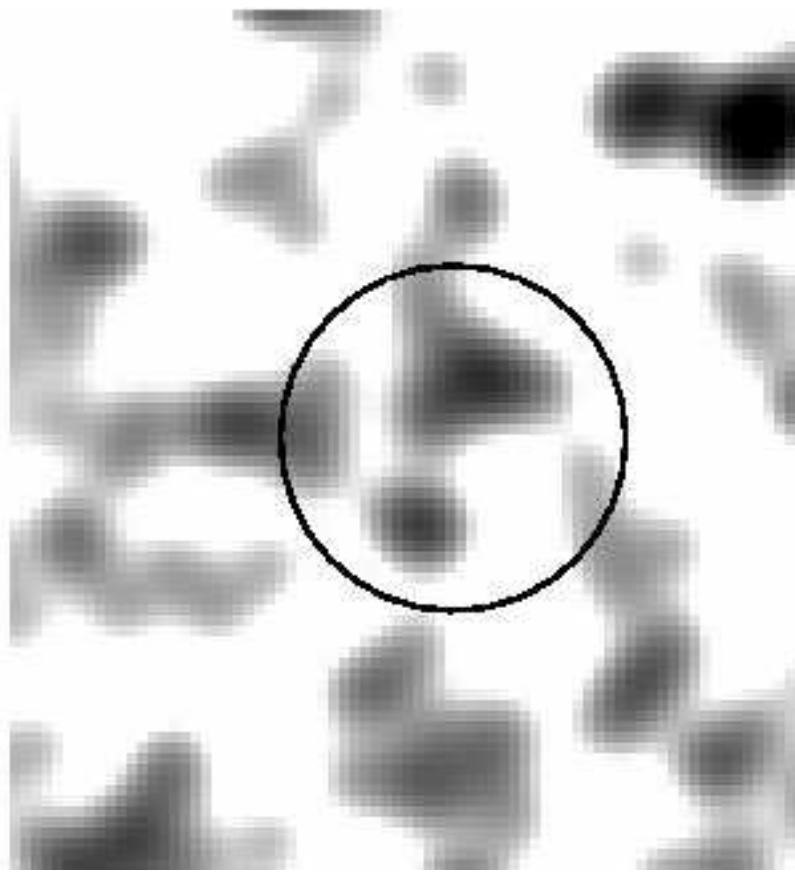
\end{document}